\documentstyle[twocolumn,prl,aps]{revtex}

\DeclareMathAlphabet{\bi}{OML}{cmm}{b}{it}

\newcommand{\p}{\partial}

\begin{document}
\draft

\title{Semi-classical magnetoresistance in weakly modulated magnetic fields}

\author{A. Matulis \cite {*:gnu} and F. M. Peeters \cite {f:gnu}}
\address{Department of Physics, University of Antwerp (UIA), B-2610
Antwerp, Belgium}

\date{\today}
\maketitle

\begin{abstract}
The semi-classical conductance of a two-dimensional electron gas is calculated
in the presence of a one-dimensional modulated magnetic
field with zero average. In the limit of small magnetic field amplitudes (B)
the contribution of the magnetic modulation to the magnetoresistance increases as
$B^{3/2}$ in the diffusive limit, while the increase is linear in $B$ in the
ballistic regime. Temperature does not influence the power law behavior
but it decreases the prefactor of this functional behavior.
\end{abstract}

In recent years there has been an increased interest\cite{peeters99}
in hybrid systems which
promise to increase the functionality of present day semiconductor devices.
One example of such type of systems are those in which 
semiconductors and magnetic materials are combined where the magnetic material
provides a local magnetic field which
influences locally the motion of the electrons in the semiconductor. The latter
is usually a heterostructure which contains a two-dimensional electron gas
(2DEG). The 2DEG acts as a detector measuring the magnetic state of the
magnetic material.
Previously, the coupling between such a non homogeneous magnetic field and the
2DEG was demonstrated \cite{carmona} in the case the magnetic field ($B$) was
directed perpendicular to the 2D plane. 
In this case one has a modulation of the $B$-field on top of a homogeneous
background field and the influence of the $B$-field modulation on the
2DEG is relatively weak \cite{peeters92}.

When the magnetic field is directed parallel to the 2DEG the magnetic material
is magnetized parallel to the 2D plane which leads to fringing fields near the
edge of the magnetic material having a non zero magnetic field component
perpendicular to the 2D layer. Those fields form a magnetic
barrier\cite{peeters93,matulis94} for the electron motion in the 2D plane.
Because now there is no background perpendicular magnetic field for the 2DEG
the influence on the resistance of such magnetic barriers is much more
pronounced and large increases in the magnetoresistance have been found
\cite{kubrak,ihn,kato}.

In the present work we investigate the magnetotransport in weak modulated
magnetic fields for which the average $B$-field is zero. For the case the typical
magnetic energy is much smaller than the Fermi energy a semi-classical
analysis is applicable.
We find that in the diffusive regime
the correction to the magnetoresistance exhibits a {\it non analytical}
behavior in the limit of small magnetic field amplitudes which differs 
from the behaviour in the ballistic regime. 

We consider electrons moving in a two-dimensional (2D) $xy$-plane.
The magnetic field, directed along the $z$-direction,
is periodic along the $x$-direction $\vec{B} = B (0,0,b(x/l_0))$ with period
$l_0$, where $b(x)$ is a periodic ($b(x+1)=b(x)$) dimensionless function
describing the magnetic field modulation with zero average value.

In a semi-classical analysis the electron motion in a magnetic field
is described by the following Hamiltonian (or its energy):
\begin{equation}\label{energy}
  \varepsilon = \frac{1}{2m}\left\{p_x^2 + \left[p_y - \frac{eBl_0}{c}a(x/l_0)
  \right]^2 \right\} ,
\end{equation}
where $m$ is the electron effective mass, $\vec{p}=(p_x,p_y)$ is
the electron canonical momentum, and the dimensionless periodic
function
\begin{equation}
  a(x) = \int^x dx' b(x') ,
\end{equation}
characterizes the vector potential $\vec{A} = Bl_0(0,a(x/l_0),0)$
which is taken in the Landau gauge.
The quantum energy spectrum of electrons in modulated magnetic fields was
studied in Refs. \cite{ibrahim,zwerschke}.

We restrict our analysis to the case where the electron
transition through a single period is ballistic, i.~e.\ the mean free
path $l=v_F\tau \gg l_0$ ($v_F$ is the electron Fermi velocity, and
$\tau$ the relaxation time) and the motion in the sample is diffusive,
i.e. the mean free path is
smaller than the size of the sample.
($l \ll L_x, L_y$). For diffusive transport and in the limit of 
small magnetic fields ($\omega_c\tau \ll 1$) the
expression for the average conductivity tensor is given by
the following integral over the electron phase space
$(x,p_x,p_y)$
\begin{eqnarray}\label{cond0}
  \sigma_{ij} = -\frac{e^2}{(2\pi\hbar)^2L_x}\int_0^{L_x}dx
  \int_{-\infty}^{\infty}dp_x \int_{-\infty}^{\infty}dp_y \,
  \tau v_iv_j \frac{\p f_F}{\p\varepsilon} ,
\end{eqnarray}
where the symbol $f_F(\varepsilon) = \{\exp\{(\mu-\varepsilon)/kT\}+1\}^{-1}$
stands for the equilibrium electron Fermi-Dirac distribution function
with $T$ the temperature and $\mu$ the chemical potential which equals the
Fermi energy $E_F=\pi \hbar^2 n/m$ in the zero temperature limit, where $n$ is
the 2D electron density. Note that the coordinate $y$ is excluded
from the phase space as the system is homogeneous in that direction.
The expressions for the electron velocities follow from the
Hamilton equations of motion
\begin{eqnarray}\label{velo}
  v_x &=& \frac{\partial\varepsilon}{\partial p_x}
  = \frac{1}{m}p_x , \nonumber \\
  v_y &=& \frac{\partial\varepsilon}{\partial p_x}
  = \frac{1}{m} \left\{p_y - \frac{eBl_0}{c}a(x/l_0)\right\}.
\end{eqnarray}

First, let us consider the conductivity in the 
zero temperature limit when the derivative of the Fermi function
reduces to a $\delta$-function. The component $\sigma_{yy}$ can
be calculated straightforwardly, as the sample is homogeneuos 
along the $y$-direction all trajectories have to be taken into
account. Inserting expression (\ref{velo}) into the 
conductivity expression (\ref{cond0}) we obtain
\begin{eqnarray}\label{syy}
  \sigma_{yy} &=& \frac{e^2}{(2\pi\hbar)^2L_x}\int_0^{L_x} dx
  \int_{-\infty}^{\infty} dp_x \int_{-\infty}^{\infty} dp_y \,
  \tau \frac{1}{m^2} \nonumber \\
&& \times \left\{p_y -\frac{eBl_0}{c}a(x/l_0)\right\}^2  \nonumber \\
&& \times \delta\Bigg\{E_F - \frac{1}{2m}\left\{p_x^2+ \left(p_y -\frac{eBl_0}{c}
  a(x/l_0)\right)^2\right\} \Bigg\} , 
\end{eqnarray}
which leads to the well-known magnetic field independent result
\begin{equation}\label{sigyy}
  \sigma_{yy} = \sigma_0 = \frac{e^2 E_F \tau}{2\pi\hbar^2}.
\end{equation}
We assumed that the relaxation time depends only on the electron
energy and therefore for the case of zero temperature it can be replaced
by the constant value $\tau=\tau(E_F)$.
Thus the weak magnetic field modulation in the $x$-direction does not
change the conductivity in the $y$-direction, as expected.

The calculation of $\sigma_{xx}$ is more complicated because
for sufficiently strong magnetic fields (or small electron velocity, but
such that $\hbar \omega_c \ll E_F$)
some electrons can be forced into \textit{snake orbits}.
The electron motion on such orbits oscillates in the
$x$-direction around the average value $x_0$. Therefore, such electrons
do not contribute to $j_x$, and consequently,
in the expression for the conductivity $\sigma_{xx}$ those snake
orbits have to be excluded. The classification of all possible
electron orbits are given in Fig.~\ref{fig1} where the Fermi
surface $\varepsilon(p_x,p_y,x)=E_F$ is plotted. In the case of
zero temperature only electrons with trajectories on the
Fermi surface contribute to the conductivity integral. As the
energy (\ref{energy}) does not depend on $y$, the momentum
$p_y$ is conserved, and consequently the trajectories are defined by the 
intersection of the above Fermi surface with the $p_y=C^{te}$
planes. It is apparent from
Fig.~\ref{fig1} that there are two types of trajectories. The
trajectories as indicated by symbol $D$ are able to run along the whole
$x$-axis. Such electrons are moving along \textit{open}
trajectories and they will contribute to the current along the $x$-direction. 
The other
trajectories, indicated by symbol $E$, are {\it closed} and they correspond to 
the snake orbits. Thus we can separate the conductivity into two parts 
\begin{equation}\label{sxx}
  \sigma_{xx} = \sigma_0 - \sigma_{xx}^{\mathrm{s.o.}} \ ,
\end{equation}
where the symbol $\sigma_{xx}^{\mathrm{s.o.}}$ stands for the snake
orbit contribution. The latter term defines the decrement of the
conductivity due to the modulated magnetic field,
and in the limiting case of small magnetic fields
it is proportional to the increase of the magnetoresistance due to the magnetic
field modulation. 

The snake orbits are located
above the plane $A$ and below the plane $B$. Those planes are defined by
$p_y (A) = \sqrt{2mE_F} + (eBl_0/c)a_{\mathrm{min}}$, and
$p_y (B) = -\sqrt{2mE_F} + (eBl_0/c)a_{\mathrm{max}}$,
with $a_{\mathrm{min}} = \mathrm{min}\{a(x)\}$ and $a_{\mathrm{max}} =
\mathrm{max}\{a(x)\}$ the extremal points of the vector potential $a(x)$.
Thus we can write
\begin{equation}\label{soc}
  \sigma_{xx}^{\mathrm{s.o.}} = \sigma_A + \sigma_B \ ,
\end{equation}
where the contribution from the snake orbits above the $A$ plane is given by
\begin{eqnarray}\label{snakecontr}
 \sigma_A = && \frac{e^2}{(2\pi m\hbar)^2L_x}\int_0^{L_x} dx
  \int_{-\infty}^{\infty} dp_x p_x^2 \int_{-\infty}^{\infty} dp_y \,
  \tau \nonumber \\
&& \times \delta\left\{E_F - \frac{1}{2m}\left[p_x^2+ \left(p_y -\frac{eBl_0}{c}
  a(x/l_0)\right)^2\right] \right\} \nonumber \\
&& \times \Theta\Bigg\{p_y -\frac{eBl_0}{c}a_{\mathrm{min}} - \sqrt{2\pi m\mu}\Bigg\} ,
\nonumber \\
&& = \frac{e^2 E_F \tau}{2\pi^2\hbar^2L_x}\int_0^{L_x} dx
  \int_{-\varphi_0}^{\varphi_0}\sin^2(\varphi)d\varphi.
\end{eqnarray}
Here the angular integration has to be performed over the white 
sections of the circle C in Fig.~\ref{fig1}, or equivalently, along the 
bold arch shown in Fig.~\ref{fig2}.
The limiting angle is defined as
%
$\varphi_0 = \arccos (1- \Delta p_y/\sqrt{2mE_F})$, with 
$\quad \Delta p_y = (eBl_0/c)\left\{a(x/l_0)-a_{\mathrm{min}} \right\}$.
%
In the asymptotic case of small magnetic field modulations it becomes 
$\varphi_0 = \sqrt{2\Delta p_y/\sqrt{2mE_F}}$.
In the latter case the integration in expression (\ref{snakecontr}) leads to
\begin{eqnarray}\label{rez}
  \sigma_A = \frac{e^2 E_F\tau}{2\pi^2\hbar^2L_x}\int_0^{L_x} dx
  \int_{-\varphi_0}^{\varphi_0}\varphi^2d\varphi
  = \sigma_{yy}c_A\left(\frac{B}{B_0}\right)^{3/2},
\end{eqnarray}
where $B_0 = (2\pi c\sqrt{mE_F})/(el_0)$,
and
\begin{equation}
  c_A = \frac{8\cdot 2^{1/4}\sqrt{\pi}}{3}\int_0^1 dx \{a(x)-
  a_{\mathrm{min}}\}^{3/2}.
\end{equation}
The integration over the snake orbits below the plane $B$ leads to
the same expression (\ref{rez}) except that now the coefficient $c_A$ has
to be replaced by
\begin{equation}
  c_B = \frac{8\cdot 2^{1/4}\sqrt{\pi}}{3}\int_0^1 dx \{
  a_{\mathrm{max}}-a(x)\}^{3/2}.
\end{equation}
Thus the resistance change due to the modulated magnetic field becomes
$\Delta R_{xx}/R_0 = (c_A + c_B) (B/B_0)^{3/2}$ where $R_0$ is the
resistance in the absence of any magnetic field modulation.

As a special case let us consider a simple cosine periodic magnetic field
modulation (with period $l_0$) $b(x) = \cos(2\pi x)$, which leads to
$a(x)=\sin(2\pi x)/2\pi$. In this case the coefficients can be
easily evaluated
\begin{eqnarray} 
  c_0 = c_A + c_B &&= \frac{8\cdot 2^{3/4}}{3\pi}\int_{-1/4}^{1/4}
  dx \{\sin(2\pi x)+1\}^{3/2} \nonumber \\
&& = \frac{64\cdot 2^{1/4}}{9\pi^2} \approx 0.86.
\end{eqnarray}

In the present classical ballistic situation an electron which passed through the first
magnetic barrier will also pass through the other barriers. As a consequence
the above result can also be applied to the {\it one barrier} situation.
The periodic oscillating Fermi surface shown in Fig.~\ref{fig1}
reduces now to a single step. The
trajectories of $D$-type (i.e. the open orbits) contribute to the $\sigma_{xx}$
conductivity, but the snake orbits (trajectories of $A$-type) are replaced
by trajectories which reflect from the barrier. 
The integral in expression (\ref{rez}) must now 
be evaluated over those reflected trajectories. When the barrier thickness 
$l_o\ll L_x$, the integral $\int^{L_x/2}_{-L_x/2} dx
(a(x/l_0)-a_{\mathrm{min}})^{3/2}$ becomes 
the sample length multiplied by the
total vector potential increment over the barrier: 
$L_x\{a_{\mathrm{max}}- a_{\mathrm{min}}\}^{3/2}$,
where $a_{\mathrm{max}}=a(L_x/2)$ and $a_{\mathrm{min}}=a(-L_x/2)$.

A single magnetic field barrier in a 2DEG, created experimentally by parallel
magnetization of magnetic strips placed on top of the 2DEG, can be
represented by \cite{ihn,jonas}  
\begin{equation}\label{barrier}
B_z(x) = \frac{\mu_0 M}{4 \pi} \ln \frac{x^2 + d^2}{x^2 + (d+D)^2} . 
\end{equation}
The magnetic strip has a thickness $D$
with magnetization $M$ and is placed a distance $d$ from
the 2D electron system. The vector potential is obtained by integrating (14)
over $(-L_x/2,L_x/2)$ which gives $A_{\mathrm{max}}- A_{\mathrm{min}}=\mu_0
MD/2$ where use has been made of $L_x \gg d, D$ which is valid
in typical experimental situations and $A=Bl_0a$.
Finally, we obtain for the contribution of the reflected trajectories to the 
conductivity
\begin{equation}\label{bdif}
-\frac{\sigma_{xx}^{\mathrm{r.t.}}}{\sigma_{yy}}
=\frac{\Delta R_{xx}}{R_0} = \frac{2^{1/4}}{3\pi}\left(
\frac{e\mu_0MD}{c\sqrt{mE_F}}\right)^{3/2}.
\end{equation}

%

The main feature of the obtained magnitoresistance is its non analytical
behavior $B^{3/2}$ for small magnetic field amplitudes. It is remarkable
that it does not depend on the actual form of the modulating field, but it
is determined by the density of snake orbits (or reflected trajectories)
at the Fermi surface.

For non zero temperature, the previous effect
will be suppressed by thermal fluctuations.
To generalize our results to non zero temperature
we have to replace any function
$G(E_F)$ which depends on the Fermi energy $E_F$ by the corresponding
average over the derivative of the Fermi function
\begin{equation}
  G(\mu) = \frac{1}{T} \int_0^{\infty} d\varepsilon
  \frac{\exp\left((\varepsilon-\mu)/k_BT\right) G(\varepsilon)}
  {\left\{\exp\left((\varepsilon-\mu)/k_BT\right)+1\right\}^2} .
\end{equation}
Consequently, taking into account $\sigma_{yy}$ (\ref{sigyy}) and the 
dependence of $B_0$ on $E_F$ we obtain for the 
snake orbit contribution to the conductivity  
\begin{eqnarray}\label{tint}
 \sigma_{xx}^{\mathrm{s.o.}}(T)
  = && \left(\frac{e^2\tau c_0}{2\pi\hbar^2}\right)
  \left( \frac{Bel_0}{c\sqrt{m}} \right)^{3/2}
  \frac{1}{T} \nonumber \\
&& \times \int_0^{\infty}d\varepsilon \frac{\exp\left(
  (\varepsilon-\mu)/k_B T\right) \varepsilon^{1/4}}
  {\left\{\exp\left((\varepsilon-\mu)/k_B T\right)+1\right\}^2},
\end{eqnarray}
where we assumed that the relaxation time does not depend on the energy.
In the limit of small temperatures the integral
can be evaluated analytically 
%
and we arrive at the final expression for the conductivity along
the direction of the magnetic field
modulation
\begin{equation}
  \frac{\sigma_{xx}}{\sigma_0} =
  \left\{1 - c_0(T)\left(\frac{B}{B_0}\right)^{3/2} \right\} ,
\end{equation}
where
\begin{equation}
c_0(T) = c_0 \left\{1-\frac{\pi^2}{32}\left(\frac{k_B T}{\mu}\right)^2 \right\}.
\end{equation}
Notice that temperature decreases the nonanaliticity coefficient $c_0$
(with about 30\% when $k_B T=\mu$) but does not influence the power law dependence.

Tunneling through magnetic barriers in the ballistic regime was studied
numerically in Refs. \cite{peeters93,matulis94}.
For weak barrier only electrons impinging on the magnetic barrier under 
an angle $\theta
\approx \pi/2$ (i.e. $\cos(\theta) \approx (\pi/2 - \theta) = \phi$)
are reflected and thus we obtain from Eq. (6) of Ref.
\cite{matulis94} the conductance change due to tunneling through a weak
magnetic barrier 
\begin{equation}
-\Delta G \approx G_0 \int_0^{\varphi_0} \varphi d\varphi ,
\end{equation}
where $G_0=e^2mv_FL_y/\hbar^2$.
Notice that the difference with the diffusive case is
the power of the angle $\varphi$ in the expression for the
conductance/conductivity. In the case of ballistic transport
the conductance is proportional to $v_x$, while for diffusive
transport, see Eq. (\ref{cond0}), we have $v_x^2$. This difference 
leads to a linear $B$-dependence in the ballistic regime 
\begin{equation}\label{bbal}
-\frac{\Delta G}{G_0} = \frac{\Delta R_{xx}}{R_0}
= \frac{\Delta p_y}{\sqrt{2mE_F}} =
\frac{e\mu_0 MD}{2c\sqrt{2mE_F}}.
\end{equation}
The above expression is obtained for the single barrier (\ref{barrier}) and 
agrees with Eq.~(3) of Ref. \cite{ihn}.

Lets compare these results with the experiments of Refs. \cite{kubrak,ihn}.
In Ref. \cite{kubrak} a Co strip of thickness $D=90 nm$ was placed a distance
$d=35 nm$ from a 2DEG formed in a
GaAs-heterostructure with $E_F = 15.7 meV$. The magnetization of the Co strip
\cite{kubrakp} was $\mu_0 M \approx 9 B$ with a saturation magnetization of
$1.6 T$. In this experiment one is in the diffusive regime and inserting these
values in Eq. (15) we obtain $\Delta R_{xx}/R_0 = 4.3 B(T)^{3/2}$. The
estimated zero field resistance \cite{kubrakp} 
was $R_0 = 2.35 \Omega$ which results into $\Delta
R_{xx}(\Omega ) = 10.1 B(T)^{3/2}$ and agrees with the experimental
\cite{kubrakp} low magnetic field behavior $\Delta R_{xx}(\Omega) = 9
B(T)^{3/2}$. 

In contrast, the experiments of Van\v cura {\it et al} \cite{ihn} are closer to the
ballistic regime (they have been performed on shorter samples, i.e. $L_y = 34
\mu m$). They placed rectangular Co dots of thickness $D=100 nm$ above
a 2DEG of width $L_y=W=20 nm$ in a GaAs-heterostructure with $E_F = 19.7 meV$. 
Inserting these values in Eq. (19) we obtain $\Delta R_{xx}/R_0 = 3.6 B(T)$
where $R_0 = 1/G_0 = 1.01 \Omega$. This gives $\Delta R_{xx} (\Omega ) = 3.96 B(T)$
which is a factor 2.2 smaller than the experimental result
$\Delta R_{xx} (\Omega) = 8.75 B(T)$. 
At the saturation field of $1.5 T$ we find
theoretically $\Delta R_{xx} = 5.94 \Omega$ which compares with the
experimental result $\Delta R_{xx} = 3.5 \Omega$ and which is a factor 1.7
smaller. Thus on the average we find a reasonable agreement between theory and
experiment but it is clear that all details are not yet fully understood. The
discrepancy may be due to the fact that 
the Co dots are not homogeneously magnetized. 
Nevertheless, the linear magnetic field dependence is nicely reproduced. 

In conclusion we obtained the change in the magneto-resistance due to the
presence of magnetic barriers with zero average magnetic field.
We found that for diffusive transport the magneto-resistance increases as
$c(T)B^{3/2}$ with the amplitude of the magnetic barrier ($B$) where the coefficient
$c(T)$ decreases with increasing temperature.
This result is different from the ballistic regime where the increase is 
linear in $B$.

Acknowledgments: This work is partially supported by the Flemish Science
Foundation (FWO-Vl), IMEC and IUAP-IV. One of us (FMP) is a research director 
with FWO-Vl and he acknowledges discussions with V. Kubrak, T. Ihn and B.
Gallagher.

\begin{figure}
\caption{Fermi surface (thick solid curve) in phase space. Planes $A$ and $B$
are surfaces for the constant of motion $p_y$ delimiting the open and closed
orbits. Curve $E$ ($D$) is an example of a closed (open) orbit.}
\label{fig1}
\end{figure}

\begin{figure}
\caption{Contour for angular integration in expression
(\protect\ref{snakecontr}). Line $A$ corresponds to the Fermi surface
intersection indicated by plane $A$ in Fig.~1.}
\label{fig2}
\end{figure}



\begin{thebibliography}{BM}
\bibitem[\dag ]{*:gnu} Permanent address: Semiconductor Physics
Institute, Go\v{s}tauto 11, 2600 Vilnius, Lithuania.
\bibitem[\ddag]{f:gnu} Electronic address: peeters@uia.ua.ac.be.
\bibitem{peeters99}
For a recent review see: F.M. Peeters and Jo De Boeck, in
{\it Handbook of nanostructured materials and
nanotechnology}, Edited by H.S. Nalwa, Vol. 3 (Academic Press, New York, 1999),
p. 345.
\bibitem{carmona}
H.A. Carmona, A.K. Geim, A. Nogaret, P.C. Main, T.J. Foster, M. Henini, S.P.
Beaumont, and M.G. Blamire, Phys. Rev. Lett. {\bf 74}, 3009 (1995);
P.D. Ye, D. Weiss, R.R. Gerhardts, M. Seeger, K. von Klitzing, K. Eberl, and H.
Nickel, Phys. Rev. Lett. {\bf 74}, 3013 (1995);
S. Izawa, S. Katsumoto, A. Endo, and Y. Iye, J. Phys. Soc. Jpn. {\bf 64}, 706
(1995).
\bibitem{peeters92}
F.M. Peeters and P. Vasilopoulos, Phys. Rev. B {\bf 47}, 1466 (1993).
\bibitem{peeters93}
F.M. Peeters and A. Matulis, Phys. Rev. B {\bf 48}, 15166 (1993).
\bibitem{matulis94}
A. Matulis, F.M. Peeters, and P. Vasilopoulos, Phys. Rev. Lett. {\bf 72}, 1518
(1994).
\bibitem{kubrak}
V. Kubrak, F. Rahman, B.L. Gallagher, P.C. Main, M. Henini, C.H. Marrows,
and M.A. Howson, Appl. Phys. Lett. {\bf 74}, 2507 (1999);
V. Kubrak, A.C. Neumann, B.L. Gallagher, P.C. Main, M. Henini, C.H. Marrows,
and M.A. Howson, Physica E (1999).
\bibitem{ihn}
T. Van\v cura, T. Ihn, S. Broderick, K. Ensslin, W. Wegscheider, and M. Bichler
(to be published).
\bibitem{kato}
M. Kato, A. Endo, and Y. Iye, Phys. Rev. B {\bf 58}, 4876 (1998); 
M. Kato, A. Endo, M. Sakairi, S. Katsumoto, and Y. Iye, J. Phys. Soc. Jpn. {\bf
68}, 1492 (1999); 
M. Kato, A. Endo, S. Katsumoto, and Y. Iye, J. Phys. Soc. Jpn. {\bf 68}, 2870
(1999). 
\bibitem{ibrahim}
I.S. Ibrahim and F.M. Peeters, Am. J. Phys. {\bf 63}, 171 (1995);
{\it ibid.} Phys. Rev. {\bf B 52}, 17321 (1995).
\bibitem{zwerschke}
S.D.M. Zwerschke, A. Manolescu, and R.R. Gerhardts, Phys. Rev. B {\bf 60}, 5536
(1999).
\bibitem{jonas}
J. Reijniers and F.M. Peeters, Appl. Phys. Lett. {\bf 73}, 357 (1998).
\bibitem{kubrakp} 
V. Kubrak (private communications).
\end{thebibliography}
\end{document}